\documentstyle[12pt,cmb]{article}
\input{psfig.tex}
\begin{document}
\heading{COHERENCE AND SAKHAROV OSCILLATIONS IN THE MICROWAVE
SKY\footnote{To appear in proceedings of the
XXXIst Rencontre de Moriond, `Microwave Background Anisotropies',
astro-ph/9612015, Imperial/TP/96-97/8}
}

\author{A. Albrecht } {Imperial College, London, UK.}  { $\cdot$}

\begin{abstract}{\baselineskip 0.4cm 
I discuss the origin of the ``Sakharov oscillations'' (or ``secondary
Doppler peaks'')  in standard angular
power spectra of the Cosmic Microwave Background anisotropies
calculated   for inflationary models.  I argue that these oscillations
appear because  
perturbations from inflation have a set of properties which makes them 
``passive perturbations''.  All passive perturbations undergo a period of
linear ``squeezing''  resulting in a dramatic degree of (classical) phase 
coherence of pressure waves in the photon-baryon fluid. This phase 
coherence  eventually is
reflected in  oscillatory features in the angular power spectrum of
the temperature anisotropies observed today.
Perturbations from cosmic defects are ``active perturbations'' which have
sharply contrasting  
properties.  Active perturbations are highly non-linear and the degree
of phase coherence in a given model reflects the interplay between competing
effects. A large class of active models have {\em non}-oscillatory
angular power spectra, and only an extremely exotic class has the
same degree of coherence as is found in all passive models.
I discuss the significance of the 
search for these oscillations (which transcends the testing of any
given model) and
 take a critical look at the degree to which the question of coherence
has been treated so far in the literature.
}
\end{abstract}

\section{Introduction}
The Cosmic Microwave Background (CMB) provides us with perhaps the
clearest window on the very early universe. Based just on our current
understanding, the impact of the next
generation of high resolution CMB experiments on theoretical cosmology
is guaranteed to be enormous, and full implications of the new data
have yet to be determined.

The subject of this paper is the distinctive ``Secondary Doppler
Peaks'' or ``Sakharov Oscillations\cite{sakh}'' in the angular
power spectra of the microwave anisotropies calculated for
inflationary models.  I discuss how these features reflect the very
specific properties of ``passive perturbations'' of which the
inflationary models are a subset.  The very different nature
of the defect based (or ``active'') models suppresses the tendency to
generate Sakharov oscillations, although a different mechanism {\em
can} produce Sakharov oscillations in certain active models. 

Section 2 outlines the basic ingredients of a 
of CMB anisotropy calculation, emphasizing the differences between
active and passive models.   Section 3 spells out how the Sakharov
oscillations appear in passive models.
Section 4 describes the basic properties of active models make it hard
to produce oscillations in the angular power spectrum, but also points
out how some degree of oscillation is still possible. At the end of
Section 4 I comment of the degree to which this issue has been
addressed quantitatively in the literature.
Concluding comments appear in Section 5.

Much of this paper is based on work with my collaborators
P. Ferreira, J Magueijo, and D. Coulson, as reported in
\cite{acfm,metal,fmhere}. 

\section{The evolution of the perturbations}

Most models of structure formation consider perturbation which
originate at an extremely early time (eg the GUT era or even the Planck
era) and which have very small amplitudes (of order $10^{-6}$) until
well into the matter era.
Perturbations of inflationary origin start as short wavelength
quantum fluctuations which evolve (during the inflationary period)
into classical perturbations on scales of astronomical interest.
Defect based models undergo a phase transition (typically at around
GUT temperatures, eg $T\approx 10^{16}GeV$) forming defects 
which generate inhomogeneities on all scales.  

For all these models, once the inflationary period and/or phase
transition is over, the Universe enters an epoch where all the matter
components obey linear equations except for the defects (if they are
present).  This ``Standard Big Bang'' epoch can be divided into three
distinct periods. The first of these is the ``tight coupling''
period where radiation and baryonic matter are tightly coupled and
behave as a single perfect fluid.  When the optical depth grows
sufficiently the coupling becomes imperfect and the 
``damping period'' is entered.  Finally there is the ``free
streaming'' period, where the CMB photons only interact with the
other matter via gravity.  While the second and third periods can have
a significant impact on the overall shape of the angular power
spectrum, all the physics which produces the Sakharov
oscillations takes place in the tight coupling regime, which is the
focus of the rest of this paper.

Working in in synchronous gauge, and following the conventions and
definitions in 
references \cite{vs,pst,acfm}, the Fourier space perturbation
equations are:
\begin{eqnarray}
\label{eq:1}
\dot{\tau}_{00} & = & 
\Theta_{D} + 
{1 \over 2\pi G}\left( {\dot a \over a}\right)^2
\Omega_r\dot s \left[1 + R\right] \\
\dot{\delta}_c & = & 
4 \pi G {a \over \dot a}\left(\tau_{00} - \Theta_{00}\right) 
 -{\dot a \over a}\left({3\over 2} \Omega_c + 2\left[ 1+ R \right] 
\Omega_r \right)\delta_c \nonumber
\\ & & \mbox{}- {\dot a \over a}2\left[1 + R\right]\Omega_r s\\
\label{eq:3}\ddot s &= & - { \dot R \over 1 + R}\dot s - c_s^2k^2\left(s + 
\delta_c\right) \label{eq:pert}
\end{eqnarray}
Here $\tau_{\mu\nu}$ is the pseudo-stress tensor,
 $\Theta_D \equiv
\partial_i\Theta_{0i}$, 
$\Theta_{\mu\nu}$
is the defect stress energy, $a$ is the cosmic scale factor, 
$G$ is Newton's constant, $\delta_X$ is the density contrast and
$\Omega_X$ is the mean energy density over  
critical density of
species $X$ ($X=r$ for relativistic matter, 
$c$ for cold matter, $B$ for baryonic matter), $s \equiv {3\over
4}\delta_r - \delta_c$, 
$R = \frac{3}{4}\rho_B/\rho_r$, $\rho_B$ and $\rho_r$ are
the mean densities in baryonic and relativistic matter respectively, 
$c_s$ is the speed of sound and $k$ is the 
comoving wavenumber.  The dot denotes the conformal time 
derivative $\partial_\eta$.

In the inflationary case there are no defects and $\Theta_{\mu\nu}=0$.
With suitable initial conditions these linear equations completely
describe the evolution of the perturbations.  In the defect case
$\Theta_{\mu\nu}\neq 0$, and certain components\footnote{
Conservation of stress energy allows the equations to be manipulated so
that different components of $\Theta_{\mu\nu}$ are required as input
(a matter mainly of numerical
convenience)\cite{vs,pst,durrer,ct,ngt1-2,hw,hsw}. 
}
 of
$\Theta_{\mu\nu}(\eta)$ are required as input. Cosmic defects are
``stiff'', which means $\Theta_{\mu\nu}(\eta)$ can be viewed as an
external source for these equations.  The additional equations from
which one determines $\Theta_{\mu\nu}(\eta)$ are highly non-linear,
although the solutions tend to have certain scaling properties which
allow $\Theta_{\mu\nu}(\eta)$ to be modelled using a variety of
techniques (see for example \cite{acfm,ct,durrer}).  

\section{The passive case: Squeezing and phase coherence}

Quite generically, for wavelengths larger than the Hubble radius ($R_H
\equiv a/\dot a$), Eqns [1-3] have one decaying and one
growing solution.  The growing solution reflects the
gravitational instability, and, as required of any system which
conserves phase space volume, there is a corresponding decaying
solution.  For example, in the radiation dominated epoch, the
two long wavelength solutions for the radiation perturbation
$\delta_r$ are $\delta_r \propto \eta ^{2}$ and  $\delta_r \propto
\eta ^{-2}$.  Over time, $R_H$ grows compared with a comoving
wavelength so in the Standard Big Bang epoch a given mode starts with
wavelength $\lambda >> R_H$ but eventually crosses
into the $\lambda < R_H$ regime.  In the period of tight coupling
modes with
$\lambda < R_H$ undergo oscillatory behavior since the radiation pressure stabilizes
the fluid against gravitational collapse.  This process of first
undergoing unstable behavior which eventually converts to oscillatory
behavior is the key to the formation of Sakharov oscillations.

\begin{figure}[h]
\centerline{\psfig{file=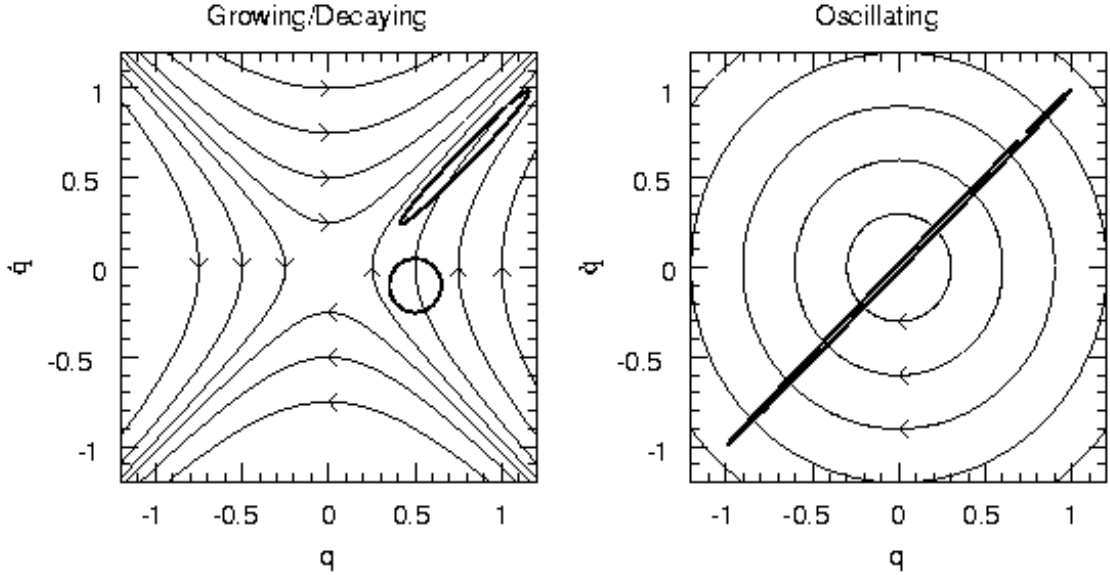,width=6in}}
\caption{Phase space trajectories for the up-side-down harmonic
oscillator (left panel) and 
right-side-up harmonic oscillator (right panel). These toy models
illustrate the growing/decaying regime and oscillating regime
(respectively).  The growing/decaying regime causes ``squeezing'',
which drives all solutions toward (and outward along) the $q=\ddot q$
axis. In a passive model the perturbations first encounter
the squeezing regime and thus the phase space
distribution which enters oscillatory regime is highly squeezed and
a unique temporal phase (up to a shift by $\pi$) is specified for the
oscillatory regime.  The elongated curve on the right panel is the
result of squeezing a circle (centered at the origin) by a factor of
$100$. Typically inflation models will have squeeze factors of $10^{20}$
or greater.}
\label{fig:squeeze}
\end{figure}

I will now illustrate the process with a simple toy model.  The
simplest example of growing/decaying behavior is given by the upside
down harmonic oscillator ($\ddot q = q$) which has the general
solution $q = Ae^t + Be^{-t}$).  Figure \ref{fig:squeeze} (left panel) shows the trajectories in
phase space for this system.  
As the ``particle'' rolls down the
inverted parabolic potential both $q$ and $\dot q$ increase
arbitrarily and {\em any} starting point in phase space evolves
arbitrarily close to the $\dot q = q$ axis.  Any initial region in
phase space will become squeezed against this axis and elongated along
it, as illustrated by the circle in the figure.  This
``squeezing\footnote{These dynamics are similar to those
producing ``squeezed states'' of light in quantum optics, but the
effect discussed here is completely classical. \cite{squeeze}
}'' process is generic to any equation which has one growing and one
decaying solution, although the simple shape of the squeezed region is
specific to models with linear equations.

The right-side-up harmonic oscillator ($\ddot q = -q$) serves to
illustrate the oscillating regime (when the wavelength is smaller than $R_H$).
The right hand panel in Fig \ref{fig:squeeze} shows the phase space
trajectories for the right-side-up harmonic oscillator.  The familiar
oscillatory behavior describes circles in the $q-\dot q$ plane (in
polar coordinates the angle corresponds to phase of the oscillation).
The linearity ensures that the period of rotation is the same on all
trajectories, thus preserving the shapes of any initial region as it
rotates around.

The physical system in question (e.g. $\delta_r$) undergoes first
squeezing and then oscillatory behavior.  During the unstable period 
the initial phase space
distribution (dictated, in the case of an inflationary scenario, by
the quantum zero-point fluctuations) is squeezed by many orders
of magnitude.  The distribution which enters the oscillatory period is
thus highly squeezed (much more so than the oblong shape depicted in
the right panel of Fig \ref{fig:squeeze}).  The end result it that the
temporal phase of oscillation is rigorously dictated by the period of
squeezing which went before.

\begin{figure}[h]
\centerline{\psfig{file=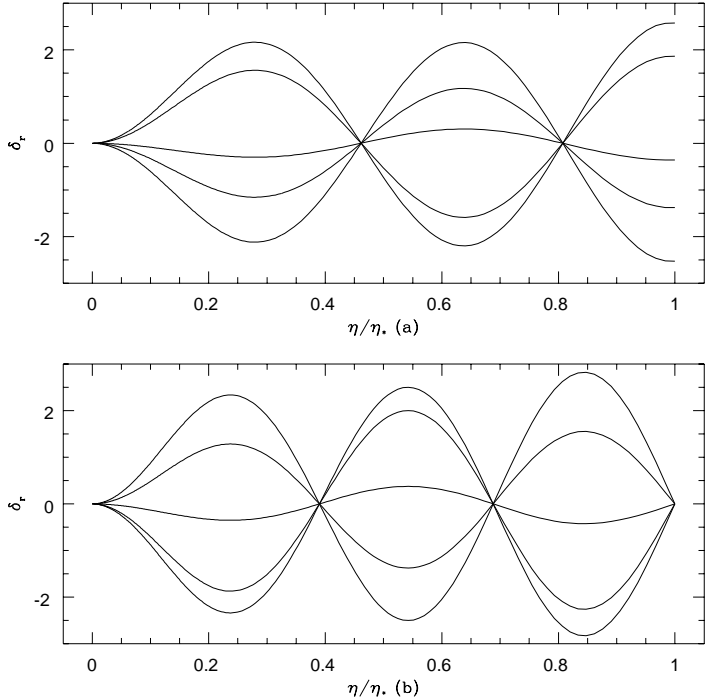,width=4in}}
\caption{Passive perturbations: Evolution of two different
 modes during the tight 
coupling era. While
in (a) elements of the ensemble have non-zero values at $\eta_\star$,
in (b), {\it all} members of the ensemble will go to zero at
the final time ($\eta_\star$), due to the fixed phase of
oscillation set by the domination of the growing solution (or squeezing) 
which occurs before the onset of the oscillatory phase.  The
y-axis is in arbitrary units.
}
\label{fig:phfo}
\end{figure}

This effect is illustrated more directly in Fig \ref{fig:phfo}, where
different solutions for $\delta_r$ are shown.  
Representative solutions are shown from across the entire ensemble,
the growing solution domination guarantees that each one goes through
zero at essentially the same time.

\begin{figure}[t]
\centerline{\psfig{file=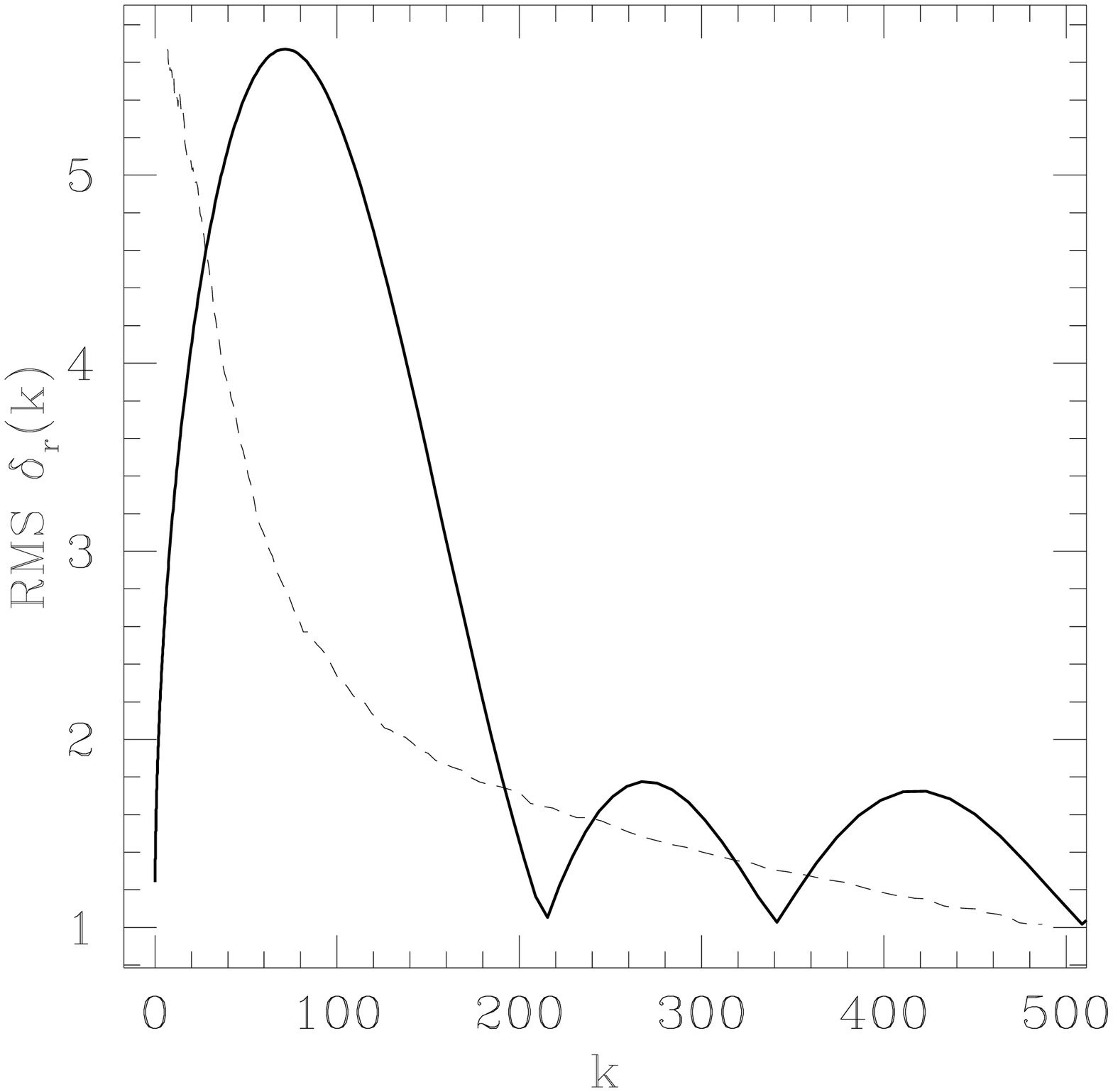,width=3in}}
\caption{The r.m.s.\ value of $\delta_r$ evaluated at decoupling
($\eta_\star$) for a passive model (solid) and an active model
(dashed). This is Fig 2 from ref [2]
where the details are presented.}
\label{fig:acfm2}
\end{figure}

Figure \ref{fig:acfm2} shows the ensemble averaged values of $\delta$
at a fixed time as a function of wavenumber (the power spectrum).
The
zeros correspond to modes which have been caught at the ``zero-point'' of
their oscillations.  The phase focusing across the entire ensemble
guarantees that there will always be some wavenumbers where the power
is zero.  It is these zeros which are at the root of the oscillations
in the angular power spectrum.

An important point is that these zeros are an absolutely fundamental
feature in any passive theory.  No amount of tinkering with the
details can counteract that fact that an extended period of liner
evolution will lead to growing mode domination, which in turn fixes
the phase of the oscillations in the tight coupling era.   If one were
to require oscillation in a passive model to be out of phase from the
prescribed value, one would imply domination by the decaying mode
outside the Hubble radius -- in other words a Universe which is not at
all Robertson-Walker on scales greater than $R_H$.  

\section{The active case: Coherence lost}

\subsection{The nature of the ensemble}

As described in Section 2, the active case is very different from the
passive case, due to the presence of what is effectively a source term
in Eqns  [1-3].  One consequence is that the whole notion of the
ensemble average is changed.  In the passive case any model with 
Gaussian initial conditions can be solved by solving Eqns
[1-3] with the initial values for all quantities given by their
initial RMS values.  The properties of linearly  evolved
Gaussian distributions guarantee that this solution will always give
the RMS values at any time.  Thus the entire ensemble is represented
by one solution.

In the active case this is not in general possible.  One has to
average over an ensemble of possible source histories, which is a much
more involved calculation.  In \cite{metal} we ``square'' the
evolution equations to write the power spectrum as convolution of
two-point functions of the sources, but there the added
complexity requires the use of the full {\em un}equal time correlation
functions.

\subsection{Non-coherence}

\begin{figure}[t]
\centerline{\psfig{file=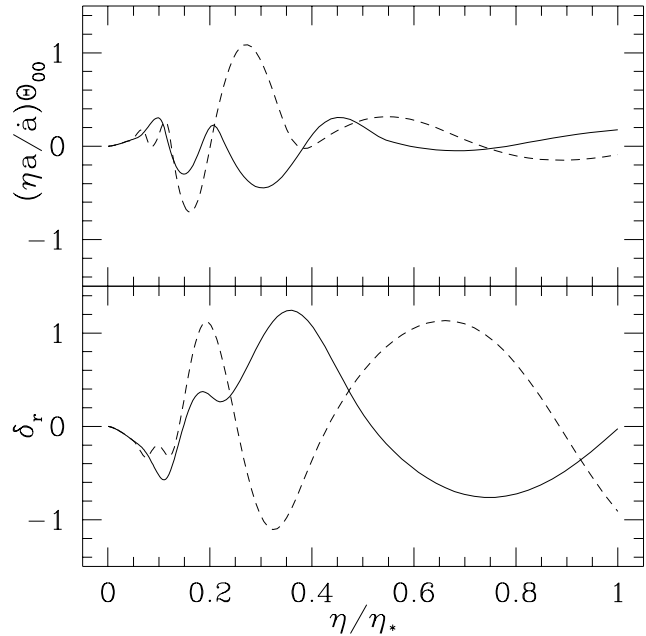,width=3.5in}}
\caption{Active perturbations:
Evolution of $\delta_r(k)$ and the corresponding source
${\Theta}_{00}$ during 
the tight  
coupling era
($\Theta_D$ is not shown).
Two members of the ensemble are shown, with matching
line types.  Due to the randomness of the source, the ensemble
includes solutions with a wide range of values at $\eta_\star$.  Unlike the
inflationary case (Figure 1) the phase of
the temporal oscillations is not fixed.  The y axis is in
arbitrary units, and the source models are the same as for figure 2.
The factor $\eta a/\dot a $ allows one to judge the relative importance (over
time) of the $\Theta_{00}$ term in Eqn 2.}  
\label{fig:acfm3}
\end{figure}

In general, the source term will ``drive'' the other matter
components, and temporal phase coherence will be only as strong as it
is within the ensemble of source terms.  An illustration of this
appears in Fig \ref{fig:acfm3}.
In many active models the sources are sufficiently decoherent that no
oscillations appear in the power spectrum (see for example the dashed
curve in Fig \ref{fig:acfm2}). 

\subsection{Coherence regained}

The source evolution is a highly non-linear process, so from the
point view of a single wavenumber the source may be viewed as a
``random'' force term.  At first glance it may seem impossible for
such random force term could induce {\em any} temporal coherence, but
here is how temporal coherence can occur:  The source term
only plays a significant role in Eqns [1-3] for a {\em finite}
period of time. This is somewhat apparent in Fig \ref{fig:acfm3}. (The
y-axis of Fig \ref{fig:acfm3} shows a quantity specially chosen to
indicate the significance of the source in Eqns [1-3].)  In
the limit where this period of significance is short compared to the
natural oscillation time of $\delta_r$ (and happens at the same time
across the entire ensemble) the ensemble of source histories {\em can}
be phase coherent.  The tendency for a given active model to produce
oscillations in the power spectrum depends on the how ``sharply
peaked'' the  significance of the source term is in time.

{\em Note added:}It is also the case that on scales larger than $R_H$
there are 
``squeezing'' mechanisms at work, even for active perturbations.  (The
gravitational instability is, after all, still present.)  In \cite{ar}
a Green function method is developed which clearly illustrates how the
active case involves a competition between squeezing effects, which
tend to produce oscillations in the power spectrum, and the
randomizing effect of the nonlinear source evolution.
In extreme cases, where the ``randomizing'' effects are minimized it
is even possible to have active models with mimic an inflationary
signal\cite{ngt1-2,hw,hsw}.

\subsection{Current Status}
So far, we are just beginning to learn the degree of coherence which
is present in the most popular active models.  Much of the work makes
use of the ``high coherence limit'' in which a single solution to Eqns
[1-3] is meant to represent the RMS value.  This only makes
physical sense for the ``sharply peaked'' sources discussed in the
previous subsection\cite{metal}, but allows one to use code designed
for passive perturbations with only minor changes.  

The question of coherence has been most aggressively pursued in the
cosmic string case\cite{acfm,metal,fmhere,hhere}, and  every
indication is  that cosmic strings give a highly decoherent ensemble.
However, there will still be room for some degree of 
skepticism  until the production of gravity waves and small loops is
incorporated in some realistic way (\cite{ra}, see also the methods
of\cite{turoksim}).  Most of the work on
cosmic texture models has not dealt quantitatively with the
question of evidence for coherence in the microwave sky although it is
pretty clear that textures should have more coherence than the cosmic
strings.  In 
\cite{ct} the ``high 
coherence'' limit was used for calculating the microwave sky, but
numerical simulations were used to illustrate some coherent behaviour
in the tight coupling era.  In other calculations the high coherence
limit has been used for convenience, and these papers have simply not
claimed to treat the question of coherence.  Recent simulations by Turok
\cite{turoksim} are the sort which can in principle address the
coherence question for a large number of active models, but the author
is not willing to claim conclusive results until a large dynamic range
is achieved.

\section{Conclusions}

The Sakharov oscillations  (or secondary Doppler peaks) in the angular
power spectrum of CMB anisotropies signify a high degree of coherence
in the primordial perturbations.  These oscillations are a certain
prediction of all passive models (which includes all inflation based
models) and can not be adjusted away\footnote{I suppose it might be possible to
construct some undulating inflaton potential which gives a
non-oscillatory temperature anisotropy spectrum, but then the oscillations would turn up
in other observations (such as those of the density field or CMB
polarization).
}.  
As such, this prediction represents probably the most clear-cut
opportunity to falsify {\em all} scenarios in which the perturbations
have an 
inflationary origin. On the other hand the observation of substantial
Sakharov oscillations in the data would have an enormous impact on the active
models, ruling out all be a very special subset of these.
 I conclude
that the search for Sakharov oscillations in the CMB sky represents an
opportunity to gain very deep insights into the origin of the
primordial perturbations.  Every effort should be made to make sure
that the experiments are able to achieve conclusive results on this matter\cite{obs}.

\acknowledgements{Special thanks to the organizers for a very productive
and enjoyable meeting.  I also wish to thank P. Ferreira, J. Magueijo,
and D. Coulson for a great collaboration, and also N. Turok
for stimulating conversations on the material presented here.  This
work was supported by the UK Particle Physics and Astronomy Council.}

\vfill
\end{document}